\documentclass[twocolumn,showpacs,preprintnumbers,amsmath,amssymb]{revtex4}
\usepackage{color}
\usepackage{bm, graphicx, amsmath}
\usepackage{hyperref}

\begin{document}

\title{Sequential discrimination of qudits by multiple observers}
\author{Mark Hillery$^{1,2}$ and Jihane Mimih$^{3}$}
\affiliation{$^{1}$Department of Physics, Hunter College of the City University of New York, 695 Park Avenue, New York, NY 10065 USA \\ $^{2}$Physics Program, Graduate Center of the City University of New York, 365 Fifth Avenue, New York, NY 10016\\ $^{3}$Naval Air Warfare Center Aircraft Division,  Patuxent River, MD 20670 USA }

\begin{abstract}
We discuss a scheme in which sequential  state-discrimination measurements  are performed on qudits to determine the quantum state in which they were initially prepared.  The qudits  belong to a set of nonorthogonal quantum  states and hence cannot be distinguished with certainty. Unambiguous state discrimination allows error-free measurements at the expense of occasionally failing to give a conclusive answer about the state of the qudit.  Qudits have the potential to carry more information per transmission than qubits.  We considered the situation in which Alice sends one of $N$ qudits, where the dimension of the qudits is also $N$.  We looked at two cases, one in which the states all have the same overlap and one in which the qudits are divided into two sets, with qudits in different sets having different overlaps.  We also studied the  robustness of our scheme against a simple eavesdropping attack and  found that by using qudits rather than qubits,  there is a greater probability that an eavesdropper will introduce errors and be detected.
\end{abstract}  


\maketitle

\section{Introduction}
It is often thought that measuring a quantum mechanical system again after it has been already measured does not yield useful information.  The first measurement places the system in an eigenstate of the observable that was measured, and a second measurement would only see this eigenstate and not the original state.  This is too narrow a view for several reasons.  First, the spectral resolution of the observable may not consist of rank-one projections, so that the same measurement result can lead to different post-measurement states.  This can also be true if the measurement is described by a POVM (positive-operator-valued measure) instead of projection operators, i.e.\ there is not a one-to-one correspondence between the measurement result and the post-measurement state.  Even in the case of when each measurement result yields a unique output state, it is possible for a second measurement to learn something about the initial state of the system \cite{rapcan}.

In an earlier paper, we studied a particular example of sequential measurements on a single system \cite{bergou1}.  A qubit in one of two states was prepared by Alice and then sent to Bob, who performed an unambiguous state discrimination measurement on it, after which Bob sent the qubit on to Charlie, who also performed an unambiguous state discrimination measurement on it.  States that are not orthogonal cannot be discriminated perfectly, and in unambiguous discrimination, while we will never make an error, the measurement can sometimes fail \cite{ivanovic,dieks,peres}.  In the sequential scenario, there is a non-zero probability that both Bob's and Charlie's measurements succeed, so that they are both able to identify the state Alice sent.  In this case Alice is able to send a bit to both Bob and Charlie using only a single qubit.

In this paper we would like to extend this scenario to higher dimensional systems, qudits, and see what effect using qudits instead of qubits has on failure probabilities and on the amount of information that can be transmitted.  In particular, Alice will send one of $N$ qudits to Bob, where the dimension of the qudits is also $N$.  We will look at two cases.  In the first, the qudit states have the property that the overlap between any two of them is the same.  In the second, the qudits are divided into two sets, and there are three possible overlaps.  Any two qudits in the same set have the same overlap, though the overlaps can be different for the two sets. Any two qudits from different sets have the same overlap, and that overlap will, in general, be different from the ones for qudits in the same set.  Using states with different overlaps could complicate the task of someone trying to obtain information about the transmissions.  In both cases, Bob will apply an unambiguous discrimination measurement to the qudit, and then send it on to Charlie.  One might think that sending one of $N$ possible qudits rather than one of two qubits might have an adverse effect on the failure probability, but we find that it does not.  Consequently, using qudits allows more information to be sent per transmission, and we will examine the channel capacity of this system.  We will also see what happens if an additional party attempts to extract information about the qudit state by using a minimum-error measurement and determine how the information thereby gained about the transmitted state depends on dimension.  These issues become important if this protocol is to be used for communication purposes.

\section{Equal overlaps}
\subsection{Qudit states with equal overlap}
Consider $N$ states in an $N$-dimensional space, $|\eta_{j}\rangle$, $j=1,2,\ldots, N$, with the property that $\langle\eta_{j}|\eta_{k}\rangle = s$ for $j\neq k$, and we shall assume that $s$ is real.  An example of such a set would be the $N$-qubit states $|\eta_{j}\rangle = |0\rangle^{\otimes (j-1)}|\mu\rangle |0\rangle^{\otimes (N-j)}$.  Here, $|\mu\rangle = \alpha |0\rangle + \beta |1\rangle$, and $s=|\alpha |^{2}$.  A second example is given by the following.  Let $\{ |j\rangle\, | j=1,2,\ldots, N\}$ be an orthonormal basis and let
\begin{equation}
|\eta_{j}\rangle = \frac{1}{[(N-1)|\alpha |^{2} +|\beta |^{2}]^{1/2}}\left( \beta |j\rangle + \alpha \sum_{k=1,k\neq j}^{N}|k\rangle\right) .
\end{equation}
In this case we have that
\begin{equation}
s=1-\frac{|\alpha - \beta |^{2}}{(N-1)|\alpha |^{2} + |\beta |^{2}} .
\end{equation}

We will assume that Alice sends one of these states, with each state being equally likely, to a second party, who then measures the state and then sends the resulting state on to the next party, who then also measures the state.  This procedure can be repeated, with each party measuring the state they receive and sending the post-measurement state on to the next party, until the qudit reaches the last party, who simply measures it.

In order to implement the procedure in the previous paragraph, we first need to find an unambiguous discrimination measurement to distinguish the states $\{ |\eta_{j}\rangle\, |\, j=1,2,\ldots, N\}$.  Unambiguous discrimination for symmetric states, of which the case of equal overlaps is an example, was studied by Chefles and Barnett \cite{chefles-barnett}, and the equal-overlap case was studied in detail by Englert and \v{R}eha\v{c}ek \cite{englert}.  Because we will need the details of the equal-overlap case in order to examine sequential measurements, we will derive this case from the beginning.   In addition, the method we use can be extended to the case when the overlaps are not equal.  The measurement operators, $\Pi_{j}$, for $j=0,1,\ldots, N$, which are positive, satisfy $\langle \eta_{j}|\Pi_{k}|\eta_{j}\rangle = 0$ for $j,k=1,2,\ldots, N$ and $j\neq k$, and $\sum_{j=0}^{N}\Pi_{j}=I$.  The operator $\Pi_{j}$, $j=1,2,\ldots, N$, corresponds to the detection of the state $|\eta_{j}\rangle$, and $\Pi_{0}$ corresponds to the measurement failing.  If we are given $|\eta_{j}\rangle$, the probability of properly identifying it is $\langle \eta_{j}|\Pi_{j}|\eta_{j}\rangle $.  Because of the condition $\langle \eta_{j}|\Pi_{k}|\eta_{j}\rangle =0$, for $1\leq j,k \leq N$ and $k\neq j$, we will never make a mistake, that is claim we found $|\eta_{k}\rangle$ when $|\eta_{j}\rangle$ was sent.  On the other hand, the measurement can fail, and this happens with a probability of $\langle \eta_{j}|\Pi_{0}|\eta_{j}\rangle$.  Clearly we have that
\begin{equation}
\label{povm}
\Pi_{j} = c_{j}|\eta^{\perp}_{j}\rangle\langle \eta_{j}^{\perp}| 
\end{equation}
where $c_{j}$ is a constant and $|\eta_{j}^{\perp}\rangle$ is the vector in the space that satisfies $\langle \eta_{k}|\eta^{\perp}_{j}\rangle = 0$ for $k\neq j$.  We will assume that all of the $c_{j}$'s are the same, that is, $c_{j}=c$.

The first thing to do is to find $|\eta_{j}^{\perp}\rangle$.  We can expand it in terms of the vectors $|\eta_{j}\rangle$,  
\begin{equation}
|\eta_{j}^{\perp}\rangle = \sum_{k=1}^{N}d_{k}|\eta_{k}\rangle .
\end{equation}
The condition $\langle\eta_{n}|\eta_{j}^{\perp}\rangle = 0$ for $n\neq j$ gives us
\begin{equation}
d_{n}=-\sum_{k\neq n} sd_{k} .
\end{equation}
Setting $\Lambda = \sum_{k=1}^{N}d_{k}$, this can be expressed as
\begin{equation}
d_{n}=\frac{-s}{1-s}\Lambda 
\end{equation}
for $n\neq j$.  Inserting this expression into the definition of $\Lambda$ we find that
\begin{equation}
\Lambda = d_{j}-(N-1)\Lambda \frac{s}{1-s} ,
\end{equation}
or
\begin{equation}
\Lambda = \frac{1-s}{1+s(N-2)} d_{j} .
\end{equation}
This implies that for $k\neq j$
\begin{equation}
d_{k}=\frac{-s}{1+s(N-2)}d_{j}
\end{equation}
and
\begin{equation}
|\eta_{j}^{\perp}\rangle = d_{j}|\eta_{j}\rangle - \frac{s}{1+s(N-2)} d_{j}\sum_{k=1,k\neq j}^{N} |\eta_{k}\rangle .
\end{equation}
One can find $d_{j}$ by normalizing the state to one, which gives
\begin{equation}
d_{j}=\left\{ \frac{1+s(N-2)}{(1-s)[1+(N-1)s]} \right\}^{1/2}
\end{equation}
and this then implies that
\begin{equation}
\label{inn-prod}
\langle\eta_{j}|\eta_{j}^{\perp}\rangle = \left\{ \frac{(1-s)[1+(N-1)s]}{1+s(N-2)} \right\}^{1/2}
\end{equation}

The next step is to find the range of $c$ that leaves $\Pi_{0}= I-\sum_{j=1}^{N} \Pi_{j}$ positive.  If we express an arbitrary vector $|\psi\rangle$ as
\begin{equation}
\label{arbvec}
|\psi\rangle = \sum_{j=1}^{N} \alpha_{j}|\eta_{j}\rangle ,
\end{equation}
the condition that $\langle\psi | (I-\sum_{j=1}^{N}\Pi_{j})|\psi\rangle \geq 0$ can be expressed as
\begin{equation}
\label{pos-matrix}
\sum_{j,k=1}^{N}\alpha^{\ast}_{j}\tilde{M}_{jk}\alpha_{k} \geq 0 ,
\end{equation}
where $\tilde{M}_{jj} = q$, $\tilde{M}_{jk}=s$ for $j\neq k$, and
\begin{equation}
\label{qc}
q = 1-c \frac{(1-s)[1+(N-1)s]}{1+(N-2)s} .
\end{equation}
We now need to find the condition on q so that $\tilde{M}$ has only non-negative eigenvalues.  Let the column vector $(x_{1}, x_{2}, \ldots, x_{N})^{T}$ be an eigenvector of $\tilde{M}$ with eigenvalue $\lambda$.  Then the eigenvalue equation for $\tilde{M}$ becomes
\begin{equation}
q x_{j} + s\sum_{k\neq j}x_{k} = \lambda x_{j} ,
\end{equation}
for $j=1,2,\ldots, N$.  Setting $\xi = \sum_{j=1}^{N}x_{j}$, we find that
\begin{equation}
(\lambda -q +s)x_{j} = s \xi .
\end{equation}
If we now sum both sides of this equation over $j$, we obtain the consistency condition
\begin{equation}
(\lambda -q + s)\xi = Ns\xi .
\end{equation}
There are now two possibilities.  If $\xi\neq 0$, then $\lambda = (N-1)s + q$.  If $\xi = 0$, then we must have $\lambda = q -s$ if at least one of the $x_{j}$ is to be nonzero.  For $\tilde{M}$ to be non-negative we need both eigenvalues to be non-negative, and this condition will be fulfilled if $q \geq s$, which implies
\begin{equation}
c\leq \frac{1+s(N-2)}{1+s(N-1)} .
\end{equation}
We can relate this condition to the overall failure probability.  If each of the $|\eta_{j}\rangle$ are equally likely, we can define the success probability of the measurement to be 
\begin{equation}
p=\frac{1}{N}\sum_{j=1}^{N}\langle\eta_{j}|\Pi_{j}|\eta_{j}\rangle = 1-q ,
\end{equation}
where the last equality follows from Eqs.\ (\ref{povm}) and (\ref{inn-prod}).  This implies that $q=1-p$ is just the failure probability of the measurement. 

\subsection{Unambiguous state discrimination using consecutive measurements}
In the previous section, we have shown that the failure probability to distinguish between $N$ qudit states cannot be lower than the overlap of the states when making an optimum single measurement. Here, we want to extend this scheme to allow for consecutive measurements. In this scenario, Alice prepares a qudit in a quantum state belonging to the set  $\{|\eta_{1}\rangle,|\eta_{2}\rangle,\ldots,|\eta_{N}\rangle\}$  and sends it to Bob. Bob performs unambiguous state discrimination on the qudit he receives and sends it on to the next party. Each party performs unambiguous state discrimination on the qudit he or she receives and sends it to the next party until all $N$ parties get the qudit and perform measurements on it.  This should be done in such a way that each party has a nonzero probability of correctly identifying the state of the qudit. The idea here is that each party, except the last, performs a non-optimal measurement on its qudit so that some information about the quantum state is left after it has been measured.  The last party, of course, can perform an optimal measurement in order to extract all of the remaining information.   The probability that Bob unambiguously detects the state $|\eta_{k}\rangle$ sent by Alice is given by
\begin{equation}
 p_{k}=\langle \eta_{k}|\Pi_{k}^{B}|\eta_{k}\rangle=c|\langle \eta_{k}|\eta_{k}^{\perp} \rangle|^{2}
\end{equation}
We need to determine the state after Bob performs his measurement since it will  become the input state for the next party's, Charlie's, measurement. This state can be expressed in terms of the detection operators $A_{k}$ which are related to the measurement operators $\Pi_{k}^{B}$ by
\begin{equation}
\Pi_{k}^{B}=A^{\dagger}_{k}A_{k}=c |\eta_{k}^{\perp} \rangle \langle\eta^{\perp}_{k}|,
\end{equation}
 for $k=1,2,\ldots,N$.  When Bob's measurement succeeds, his post-measurement state is:
\begin{equation}
|\phi_{k}\rangle =\frac{A_{k} |\eta_{k}\rangle}{||A_{k} |\eta_{k}\rangle||} ,
\end{equation}
and if Bob's measurement yields an inconclusive result, the post-measurement state is:  
\begin{equation}
|\chi_{k}\rangle = \frac{A_{0} |\eta_{k}\rangle}{||A_{0} |\eta_{k}\rangle ||} .
\end{equation} 

Since  the only requirement on $A_{k}$ is $A^{\dagger}_{k}A_{k}=c_{k} |\eta_{k}^{\perp} \rangle \langle\eta^{\perp}_{k}|$, we have considerable freedom in choosing the detection  operators $A_{k}$. We shall choose 
\begin{equation}
A_{k}=\sqrt{c} |\phi_{k}\rangle \langle\eta^{\perp}_{k}| ,
\end{equation}
where the states $|\phi_{j}\rangle$ satisfy $\langle\phi_{j}|\phi_{k}\rangle = t$, for $j\neq k$ and $t$ real.  Now when Bob's measurement succeeds, he will send a qudit in the state $|\phi_{k}\rangle$ to Charlie, and if the measurement fails, he will send a qudit in the state $|\chi_{k}\rangle$ to Charlie. For Charlie to be able to use unambiguous state discrimination, the states he wishes to distinguish need to be linearly independent \cite{Chefles}. Since we are in an $N$-dimensional Hilbert space, it is necessary to have $|\chi_{k}\rangle=|\phi_{k}\rangle$. We can therefore express $A_{0}$ as
\begin{equation}
A_{0}= \sum_{k} a_{k} |\phi_{k}\rangle \langle\eta^{\perp}_{k}|,
\end{equation} 
where the $a_{k}$ are constants that need to be determined. 
We then have 
\begin{eqnarray}
\langle \eta_{k}|A^{\dagger}_{0}A_{0}|\eta_{k}\rangle & = & |a_{k}|^{2}r=q_{k} \nonumber \\
\langle \eta_{k}|A^{\dagger}_{0}A_{0}|\eta_{m}\rangle & = & a_{m} a_{k}^{*} r  t ,
\end{eqnarray}
where $r=|\langle \eta_{j}|\eta_{j}^{\perp}\rangle |^{2}$ and $q_{k}$ is the failure probability to identify the state $|\eta_{k}\rangle$.  Due to the symmetry of the problem, we shall choose all of the coefficients $a_{k}$ to be the same, that is $a_{k}=a$ for $k=1,2,\ldots, N$, which implies that the failure probability for each of the states is the same, that is $q_{k}=r|a| = q$. The operator $A^{\dagger}_{0}A_{0}$ can now be expressed in a matrix form in the basis $\{ |\eta_{k}\rangle\, | k=1,2,\ldots, N\}$ as
\begin {equation}
A^{\dagger}_{0}A_{0}=\begin{pmatrix}
q          & qt & \cdots &  qt\\
qt        & q &  \cdots & qt\\
\vdots   &  \vdots & \ddots  &qt\\
qt      & qt   & \cdots      & q
\end{pmatrix}
\end{equation}
The operator $A^{\dagger}_{0}A_{0}$ can also be written as $A^{\dagger}_{0}A_{0}=I-c\sum_{k} |\eta_{k}^{\perp} \rangle \langle \eta_{k}^{\perp}| $, which in  matrix form can be expressed as
\begin {equation}
A^{\dagger}_{0}A_{0}=\begin{pmatrix}
1-c r          &s & \cdots &  s\\
s        &1-c r  &  \cdots & s\\
\vdots   &  \vdots & \ddots  &qt\\
s      & s   & \cdots      & 1-c r 
\end{pmatrix}
\end{equation} 
Comparing the two different  expressions for the failure operator $A^{\dagger}_{0}A_{0}$ in Eqs. (28) and (29) , we see that they will be consistent if $q=\frac{s}{t}$ and $q=1-cr$. The second condition already follows from the Eqs.\ (\ref{inn-prod}) and (\ref{qc}). The first condition and the condition $q\geq s$ from section II imply that $t \geq s$.

Let us now summarize the situation, and for the moment, assume that the only actors are Alice, Bob and Charlie.  Alice sends one of the states $\{ |\eta_{j}\rangle\, |j=1,2,\ldots, N\}$, say $|\eta_{k}\rangle$, to Bob.  Bob then measures the state.  Whether his measurement succeeds or fails, it is so designed that he sends the state $|\phi_{k}\rangle$ on to Charlie.  Note that since $t\geq s$, the states Charlie receives are, in general, less distinguishable than those in Alice's original set.  This is because Bob's measurement has extracted some information from the state.  Charlie then performs an optimal unambiguous measurement on the states $\{ |\phi_{j}\rangle\, |j=1,2,\ldots, N\}$.  Let us call Bob's failure probability $q_{B}$ and Charlie's $q_{C}$.  From our results we have that $q_{B}\geq s$, $q_{C}\geq t$, and $q_{B}=s/t$, where $t\geq s$.  Now if Charlie's measurement is optimal, he will have $q_{C}=t$, which then implies that $q_{B}q_{C}=s$.  If we want Bob and Charlie to have equal failure probabilities, we will choose $q_{B}=q_{C}=\sqrt{s}$.  This implies that the probability that both measurements succeed is $(1-\sqrt{s})^{2}$.  An identical result was found earlier for qubits \cite{bergou1}.  This shows that there is a considerable advantage to using qudits.  Since the probability for both measurements to succeed is the same for qubits and qudits, that is, it is independent of the dimension of the system, and more information can be sent using qudits, it is clearly advantageous to use higher dimensional systems.

Now let us extend the scheme to more actors.  Instead of Alice, Bob, and Charlie, we will have Alice, ${\rm Bob}_{1}$, \dots ${\rm Bob}_{M}$.  Alice sends a qudit in the state $|\eta_{k}\rangle$, which belongs to a known set $\{ |\eta_{1}\rangle, \cdots, |\eta_{N}\rangle\} $ to ${\rm Bob}_{1}$, who performs an unambiguous state discrimination measurement to  extract information about the quantum state he received. The measurement succeeds with a probability $1-q_{1}$, where $q_{1}\geq s$, and it results in a state $|\phi_{k}^{(1)}\rangle$.  The states $\{ |\phi_{k}^{(1)}\rangle\, |k=1,2,\ldots, N\}$ have an  overlap $t^{(1)}$, which exceeds the overlap of the initial sates $s$, i.e.\ $t^{(1)}\geq s$.  ${\rm Bob}_{1}$ then sends the state $|\phi_{k}^{(1)}\rangle$ on to ${\rm Bob}_{2}$.  This continues until the qudit reaches ${\rm Bob}_{M}$. While  ${\rm Bob}_{1}$ through ${\rm Bob}_{M-1}$ perform non-optimized measurements to extract information about the state, ${\rm Bob}_{M}$ performs an optimized measurement to extract all the information remaining in the quantum state since the last post-measurement state does not need to carry any further information about the initial state.  This implies that the overlap of the final set of post-measurement states, $t^{(M)}$ will be $1$.  The probability that all of the measurements succeed is
\begin{equation}
p_{succ}=\prod_{l=1}^{M} (1-q_{l}),
\end{equation}
where $q^{(l)}$ is the probability that the measurement made by ${\rm Bob}_{l}$ fails.  The total success probability is just the success probability for each input state times the probability of the corresponding input state.  Since, in our case, each state is equally likely and the success probability for each state is the same, the total success probability is the same as the success probability for each state, which is the result given in the above equation.

When ${\rm Bob}_{1}$ makes a measurement on the state he receives from Alice, his failure probability to identify the state is  $q^{(1)}=s/t^{(1)}$ where $s$ is the overlap of the initial sates while $t^{(1)}$ is the overlap of the post-measurement states.  The failure probability associated with the second measurement must satisfy a similar constraint that can be obtained by replacing $s$ by the overlap of the new input states $t^{(1)}$ and $t^{(1)}$ by $t^{(2)}$, the overlap of the post-measurement states resulting from the second measurement. The final measurement is characterized by the fact that the overlap between the post-measurement states is one since the measurement is optimum at this stage, which results in $q^{(M)}=t^{(M-1)}$. 
These constraints can be summarized in the following equations:
\begin{equation}
 \begin{matrix}
 q^{(1)} & = & s / t^{(1)} \\
q^{(2)}& = & t^{(1)} / t^{(2)} \\
& \vdots &\\
q^{(M)}& =& t^{(M-1)} / t^{(M)} = t^{(M-1)}
\end{matrix}
\end{equation}
If we assume for simplicity that $q^{(1)}=q^{(2)}=\ldots =q^{(M)}=w$, we find that $w=s^{\frac{1}{M}}$.  The success probability is then given by 
\begin{equation}
p_{succ}=(1-s^{\frac{1}{M}})^{M}
\end{equation}

\section{Qudit states with different overlaps}
Now let us consider $N$ $N$-qubit states of the form 
\begin{equation}
|\eta_{j}\rangle = \left\{ \begin{array} {cc} |0\rangle^{\otimes (j-1)} |\mu_{1}\rangle |0\rangle^{\otimes (N-j)} & 1\leq j \leq M \\ |0\rangle^{\otimes (j-1)} |\mu_{2}\rangle |0\rangle^{\otimes (N-j)} & M+1\leq j \leq N 
\end{array} \right.
\end{equation}
Suppose that $\langle 0|\mu_{1}\rangle = s_{1} >0$ and $\langle 0|\mu_{2}\rangle = s_{2} >0$.  The collection of states $\{ |\eta_{j}\rangle \, | \, j=1,2,\ldots N\}$ now has the property that $\langle \eta_{j}|\eta_{k}\rangle$ is equal to $s_{1}^{2}$ for $j,k \leq M$, $s_{2}^{2}$ for $j,k \geq M+1$, and $s_{1}s_{2}$ for either $j\leq M$ and $k\geq M+1$ or $j\geq M+1$ and $k\leq M$.

The first thing we need to do to unambiguously discriminate the states $|\eta_{j}\rangle$ is to find the states $|\eta_{j}^{\perp}\rangle$, as we did in the case of equal overlaps.  Because the details of the calculation are similar to those of the equal-overlap case, we present them, and explicit expressions for the vectors $|\eta_{j}^{\perp}\rangle$, in an appendix.  What we will need here are the quantities
$\Gamma_{1}$, which is equal to $|\langle\eta_{j}|\eta_{j}^{\perp}\rangle |^{2}$ for $j\leq M$,
\begin{equation}
\Gamma_{1}=\frac{1}{D_{1}} [(1-s_{1}^{2})(D_{1}+s_{1}^{2}(1-s_{2}^{2}))] ,
\end{equation} 
and $\Gamma_{2}$, which is equal to  $|\langle\eta_{j}|\eta_{j}^{\perp}\rangle |^{2}$ for $j \geq M+1$,
\begin{equation}
\Gamma_{2}= \frac{1}{D_{2}} [(1-s_{2}^{2})(D_{2}+s_{2}^{2}(1-s_{1}^{2}))] .
\end{equation}
See the appendix for explicit expressions for $D_{1}$ and $D_{2}$.

For the POVM elements we choose
\begin{eqnarray}
\Pi_{j} & = & c_{1} |\eta_{j}^{\perp}\rangle\langle\eta_{j}^{\perp}| \hspace{5mm} j\leq M \nonumber \\
\Pi_{j} & = & c_{2} |\eta_{j}^{\perp}\rangle\langle\eta_{j}^{\perp}| \hspace{5mm} j\geq M+1 ,
\end{eqnarray}
and $\Pi_{0}=I-\sum_{j=1}^{N}\Pi_{j}$.  The positivity condition that $\langle\psi |\Pi_{0}|\psi\rangle \geq 0$ becomes (see Eqs.\ (\ref{arbvec}) and (\ref{pos-matrix}))
\begin{equation}
\sum_{j,k=1}^{N} \alpha_{j}^{\ast}\tilde{L}_{jk}\alpha_{k} \geq 0 ,
\end{equation}
where $\tilde{L}_{jj}=1-c_{1}\Gamma_{1}$ for $j\leq M$, $\tilde{L}_{jj} = 1-c_{2}\Gamma_{2}$ for $j\geq M+1$, $\tilde{L}_{jk}=s_{1}^{2}$ for $j,k\leq M$ and $j\neq k$, $\tilde{L}_{jk}=s_{2}^{2}$ for $j,k\geq M+1$ and $j\neq k$, and finally $\tilde{L}_{jk}=s_{1}s_{2}$ for $j\leq M$ and $k\geq M+1$ or $j\geq M+1$ and $k\leq M$.  

In order to show that $\tilde{L}$ is positive, we need to find its eigenvalues.   As before, letting the eigenvector of $\tilde{L}$ be the column vector $(x_{1},x_{2}, \ldots x_{N})^{T}$, the eigenvalue equations are for $j\leq M$
\begin{equation}
(1-\Gamma_{1}c_{1})x_{j} + s_{1}^{2}\sum_{k=1,k\neq j}^{M}x_{k} +s_{1}s_{2}\sum_{k=M+1}^{N} x_{k} = \lambda x_{j} ,
\end{equation}
and for $k\geq M+1$
\begin{equation}
s_{1}s_{2}\sum_{k=1}^{M}x_{k} + (1-\Gamma_{2}c_{2})x_{j} + \sum_{k=M+1,k\neq j}^{N} s_{2}^{2} x_{k} = \lambda x_{j} .
\end{equation}
Setting
\begin{equation}
\xi_{1}= \sum_{k=1}^{M}x_{k} \hspace{5mm} \xi_{2} = \sum_{k=M+1}^{N}x_{k} ,
\end{equation}
and summing the first of the above equations from $j=1$ to $M$ and the second from $j=M+1$ to $N$, we find
\begin{eqnarray}
0 & = & [(M-1)s_{1}^{2} -\lambda +1-\Gamma_{1}c_{1}] \xi_{1}+Ms_{1}s_{2}\xi_{2} \nonumber \\
0 & = & (N-M)s_{1}s_{2}\xi_{1} \nonumber \\
& & + [(N-M-1)s_{2}^{2} -\lambda+1-\Gamma_{2}c_{2}] \xi_{2} .
\end{eqnarray}
These equations will have nonzero solutions for $\xi_{1}$ and $\xi_{2}$ if the determinant of the coefficients of the above equations vanishes.  Setting
\begin{equation}
F_{1}=1-\Gamma_{1}c_{1}-s_{1}^{2} \hspace{5mm} F_{2}=1-\Gamma_{2}c_{2}-s_{2}^{2} ,
\end{equation}
this condition becomes 
\begin{eqnarray}
0 & = & \lambda^{2} - \lambda [Ms_{1}^{2}+(N-M)s_{2}^{2}+F_{1}+F_{2}] \nonumber \\
 & & +F_{1}F_{2} + Ms_{1}^{2}F_{2} + (N-M)s_{2}^{2}F_{1} .
\end{eqnarray}
The solutions to this equation give us two of the eigenvalues of $\tilde{L}$, and they will be non-negative if
\begin{equation}
\label{A1A2}
Ms_{1}^{2}F_{2} + (N-M)s_{2}^{2}F_{1} + F_{1}F_{2} \geq 0 .
\end{equation}
The other eigenvalues come from the case $\xi_{1}=\xi_{2}=0$.  In that case the eigenvector equations become
\begin{equation}
(1-\Gamma_{1}c_{1}-s_{1}^{2}-\lambda )x_{j}=0 ,
\end{equation}
for $j\leq M$ and
\begin{equation}
(1-\Gamma_{2}c_{2}-s_{2}^{2} -\lambda )x_{j}= 0 ,
\end{equation}
for $j\geq M+1$.  In order that not all of the $x_{]}$ for $j\leq M$ be equal to zero, we need $\lambda = 1-\Gamma_{1}c_{1}-s_{1}^{2}$, and this and the condition $\xi_{1}=0$ results in $M-1$ eigenvectors.  For not all of the $x_{j}$ for $j\geq M+1$ to be zero, we need $\lambda = 1-\Gamma_{2}c_{2}-s_{2}^{2}$, and this and the condition $\xi_{2}=0$ yields $N-M-1$ eigenvectors.  Therefore, our conditions for $\tilde{L}$ to be non-negative are
\begin{equation}
\label{pos-cond}
1-s_{1}^{2}\geq \Gamma_{1}c_{1} \hspace{5mm} 1-s_{2}^{2} \geq \Gamma_{2}c_{2} .
\end{equation}
Note that these conditions are equivalent to $F_{1}\geq 0$ and $F_{2}\geq 0$, so that if they are satisfied, so is the condition in Eq.\ (\ref{A1A2}).

Next we need to specify the post-measurement states and the detection operators.  We want the state $|\eta_{j}\rangle$ to go to the state $|\phi_{j}\rangle$ after the measurement, where $\langle \phi_{j}|\phi_{k}\rangle$ is equal to $t_{1}^{2}$ for $j,k \leq M$, $t_{2}^{2}$ for $j,k \geq M+1$, and $t_{1}t_{2}$ for either $j\leq M$ and $k\geq M+1$ or $j\geq M+1$ and $k\leq M$.  This will be the case if we choose the detection operators to be
\begin{equation}
A_{j}=\left\{ \begin{array}{cc} \sqrt{c_{1}} |\phi_{j}\rangle\langle \eta_{j}^{\perp}| & 1\leq j \leq M \\
\sqrt{c_{2}}|\phi_{j}\rangle\langle \eta_{j}^{\perp}| & M+1 \leq j \leq N ,\end{array} \right. 
\end{equation}
and 
\begin{equation}
A_{0} = a_{1}\sum_{j=1}^{M}  |\phi_{j}\rangle\langle \eta_{j}^{\perp}| + a_{2} \sum_{j=M+1}^{N}  |\phi_{j}\rangle\langle \eta_{j}^{\perp}| ,
\end{equation}
with $a_{1}$ and $a_{2}$ to be determined.  Defining $q_{1}=1-c_{1}\Gamma_{1}$, which is the failure probability for the states $|\eta_{j}\rangle$, $1\leq j \leq M$, and $q_{2}=1-c_{2}\Gamma_{2}$, which is the failure probability for $|\eta_{j}\rangle$, $M+1\leq j \leq N$, and setting $A_{0}^{\dagger}A_{0}$ equal to $I-\sum_{j=1}^{N}\Pi_{j}$, we find that
\begin{eqnarray}
q_{1} & = & a_{1}^{2}\Gamma_{1} = \frac{s_{1}^{2}}{t_{1}^{2}} \nonumber \\
q_{2} & = & a_{2}^{2}\Gamma_{2} = \frac{s_{2}^{2}}{t_{2}^{2}} \nonumber \\
s_{1}s_{2} & = & a_{1}a_{2}t_{1}t_{2} \sqrt{\Gamma_{1}\Gamma_{2}} = \sqrt{q_{1}q_{2}}t_{1}t_{2} .
\end{eqnarray}
Note that the condition in the last line is consistent with the two in the first two lines.  Also note that the positivity conditions in Eq.\ (\ref{pos-cond}) can be stated as $q_{1}\geq s_{1}^{2}$ and $q_{2}\geq s_{2}^{2}$.

Now let us look at the situation where Alice sends a qudit to Bob, who measures it, and then sends it on to Charlie, who also measures it.  The failure probabilities for Bob's measurement are $q_{1B}\geq s_{1}^{2}$ and $q_{2B}\geq s_{2}^{2}$, and those for Charlie's measurement are $q_{1C}\geq t_{1}^{2}$ and $q_{2C}\geq t_{2}^{2}$.  Charlie would perform an optimal unambiguous discrimination measurement, so we would have $q_{1C}= t_{1}^{2}$ and $q_{2C}= t_{2}^{2}$.  In this case the equations in the previous paragraph imply that $q_{1B}q_{1C}=s_{1}^{2}$ and $q_{2B}q_{2C} =s_{2}^{2}$.  In the case that Bob and Charlie have the same failure probabilities, we have $q_{1B} = q_{1C}=s_{1}$ and $q_{2B} = q_{2C} =s_{2}$.

\section{Channel capacity}
Let us begin with the simplest situation, just two actors, Alice and Bob, with Alice sending qubits in one of two nonorthogonal states to Bob.  Using unambiguous discrimination, Bob either identifies the state with a probability $p$, or fails to do so with a probability $q=1-p$.  This situation is described by a binary erasure channel.  Alice sends a classical bit, which is either $0$ or $1$ to Bob, and Bob is able to read the bit with a probability $p$.  This channel can be characterized by its channel capacity.  The channel capacity for any discrete memoryless channel is defined to be the mutual information between the sender and receiver (in this case Alice and Bob) maximized over probability distributions of channel inputs.  It is the maximum rate at which information can be sent through the channel \cite{cover}.  The channel capacity for a binary erasure channel in which a fraction $q$ of the bits are erased is just $1-q$.

In order to determine how much of an advantage using qudits has over using qubits, we need to find the channel capacity of an erasure channel that has $N$ possible inputs instead of two.  Let the channel inputs be described by a random variable $X$ and the outputs be described by a random variable $Y$.  The possible inputs are $\{ x\}$, which occur with probability $P_{in}(x)$, and the possible outputs are $\{ y\}$, which occur with probability $P_{out}(y)$.  The channel is characterized by a conditional probability, $P(y|x)$, which is the probability of detecting $y$ at the output if the input was $x$.  The mutual information between the input and the output, $I(Y;X)$ is given by
\begin{equation}
I(Y;X)=H(Y)-H(Y|X) ,
\end{equation}
where $H(Y)$ is the Shannon information of $Y$,
\begin{equation}
H(Y) = -\sum_{y}P_{out}(y)\log P_{out}(y) ,
\end{equation}
and the conditional entropy is
\begin{equation}
H(Y|X)=-\sum_{x,y}P_{in}(x)P(y|x)\log P(y|x) .
\end{equation}
The logarithms here are base $2$.

In our case, both the input set is $\{ 1,2,\ldots, N\}$, and the output set is $\{ 1,2,\ldots N, e\}$, where $e$ corresponds to the case that the input is erased, or, in the quantum case, the failure of the measurement.  We then have, for the equal overlap case,
\begin{equation}
P(y|x)=\left\{ \begin{array}{cc} (1-q_{e}) & y=x \\ q_{e} & e \\ 0 & {\rm otherwise} , \end{array} \right.
\end{equation}
where $q_{e}$ is the probability that the input is erased.  This implies that
\begin{equation}
P_{out}(y)=\sum_{x}P(y|x)P_{in}(x) = \left\{ \begin{array}{cc} (1-q_{e})P_{in}(y) & y\neq e \\ q_{e} & y=e  .\end{array} \right.
\end{equation} 
We now find that
\begin{eqnarray}
\label{entropies}
H(Y|X) & = & h(q_{e})  \nonumber \\
H(Y) & = & h(q_{e}) \nonumber \\
& & - (1-q_{e}) \sum_{y\neq e}P_{in}(y)\log P_{in}(y) .
\end{eqnarray}
where 
\begin{equation}
h(q_{e}) = -q_{e}\log q_{e} -(1-q_{e})\log (1-q_{e}) .
\end{equation}
Now the channel capacity, $C$, is
\begin{equation}
C = \max_{P_{in}} I(Y;X) ,
\end{equation}
and the maximum of the second term in $H(Y)$, see Eq.\ (\ref{entropies}), occurs when $P_{in}(x)=1/N$.  This gives us
\begin{equation}
C= (1-q_{e})\log N .
\end{equation}
Therefore, the use of qudits rather than qubits results in a $\log N$ improvement in the channel capacity, since for the Alice-Bob channel, we have $q_{e}= s$, independent of the dimension.

Now let us look at the Alice-Bob-Charlie case considered in Section II.  The capacity of the Alice-Bob channel is $(1-q_{B})\log N$ and that of the Alice-Charlie channel is $(1-q_{C})\log N$, where, again, $q_{B}$ and $q_{C}$ are independent of dimension.  We see, as before, a $\log N$ improvement by using qudits.  We can also find the capacity of the channel that groups Bob and Charlie together, i.e.\ Alice is the sender, and Bob and Charlie combined constitute the receiver.  We find that the capacity of this channel is $(1-q_{B}q_{C})\log N$.  In the case in which Charlie's measurement is optimal, we saw that $q_{B}q_{C}=s$, so that in that case the channel capacity only depends on the overlap of the initial states.

The situation becomes more complicated when the overlaps, and hence the erasure probabilities, are not the same.  As before there are $N$ inputs $x\in \{ 1,2,\ldots N\}$ and $N+1$ outputs, $y\in \{ 1,2,\ldots, N,e\}$.  However, we now have that if $x\in \{ 1,2,\ldots M\}$ then the probability that $y=e$ is $q_{1}$ and the probability that $y=x$ is $1-q_{1}$.   If $x\in \{ M+1,\ldots N\}$ then the probability that $y=e$ is $q_{2}$ and the probability that $y=x$ is $1-q_{2}$.  We now need to calculate the mutual information between Alice and Bob.  Setting
\begin{equation}
p_{1}=\sum_{x=1}^{M}P_{in}(x) \hspace{5mm} p_{2}=\sum_{x=M+1}^{N} P_{in}(x), 
\end{equation}
we find that 
\begin{equation}
H(Y|X)=p_{1}h(q_{1}) + p_{2}h(q_{2}) .
\end{equation}
The output probability distribution is now
\begin{equation}
P_{out}(y) = \left\{ \begin{array} {cc} (1-q_{1})P_{in}(y) & y\in \{1,2,\ldots M\} \\ (1-q_{2})P_{in}(y) & y\in \{M+1,\ldots N\} \\ p_{1}q_{1}+p_{2}q_{2} & y=e . \end{array} \right. 
\end{equation}
For a fixed value of $p_{1}$, $H(Y)$ will be a maximum when $P_{in}(x)=p_{1}/M$ for  $x\in \{ 1,2,\ldots M\}$ and $P_{in}(x)=(1-p_{1})/(N-M)$ when $x\in \{ M+1,\ldots N \}$.  This gives us that
\begin{eqnarray}
I(Y;X) & \leq & p_{1}q_{1}\log \left[\frac{q_{1}}{p_{1}q_{1} + (1-p_{1})q_{2}}\right] \nonumber \\
& & -p_{1} (1-q_{1}) \log \left( \frac{p_{1}}{M}\right) \nonumber \\
& & + (1-p_{1})q_{2}\log\left[\frac{q_{2}}{p_{1}q_{1}+(1-p_{1}q_{2}}\right] \nonumber \\
 & & -(1-p_{1})(1-q_{2}) \log\left(\frac{1-p_{1}}{N-M}\right) ,
\end{eqnarray}
and the bound is achievable.

At this point, we need to maximize the right-hand side of the above inequality in order to find the channel capacity, and results of doing so numerically will be presented shortly.  We can also get an idea of the behavior of the right-hand side, which we shall denote by $G$, near $q_{1}=q_{2}$ by employing a series expansion.  Let $q_{1}=q+\delta q$ and $q_{2}=q-\delta q$.  When $q_{1}=q_{2}$, the maximum of $G$ is obtained when $p_{1}=M/N$ (all inputs equally likely) so we can also set $p_{1}=(M/N)+\delta p$.  Expanding $G$ up to second order in $\delta p$ and $\delta q$, we obtain
\begin{eqnarray}
G & = & (1-q)\log N + \delta q\left(1-\frac{2M}{N}\right) \log N \nonumber \\
& & + (\delta q)^{2}\frac{2M(N-M)}{qN^{2}\ln2} \nonumber \\ 
& & + 2\left( \frac{1}{\ln 2}-\log N\right) \delta p\delta q -\frac{(1-q)N^{2}}{M(N-M)\ln 2} (\delta p)^{2} .
\nonumber \\
\end{eqnarray}
This can be maximized with respect to $\delta p$, and we find
\begin{equation}
G_{max} = (1-q)\log N + \delta q \left( 1- \frac{2M}{N}\right) \log N + O(\delta q^{2}) .
\end{equation}
$G_{max}$ is, in fact, the channel capacity.  One can check that the above expression is reasonable by noting that if $\delta q>0$, which implies that the states with $1\leq j \leq M$ have a larger failure probability than those with $M+1\leq j \leq N$, then if $M>N/2$, the channel capacity is lower than its $\delta q=0$ value, because a majority of the states have a higher failure probability.

We supplement this with numerical calculations of the channel capacity.  In the first case we examine it as a function of $q_{2}$ for a fixed value of $q_{1}$ ($q_{1}=0.5$) and different values of $M$ (see Fig.\ 1).  In the second, we plot it as a function of $M$ for $q_{1}=0.5$ and different values of $q_{2}$ (see Fig.\ 2).
\begin{figure}
\label{Fig1}
\includegraphics[scale=.4]{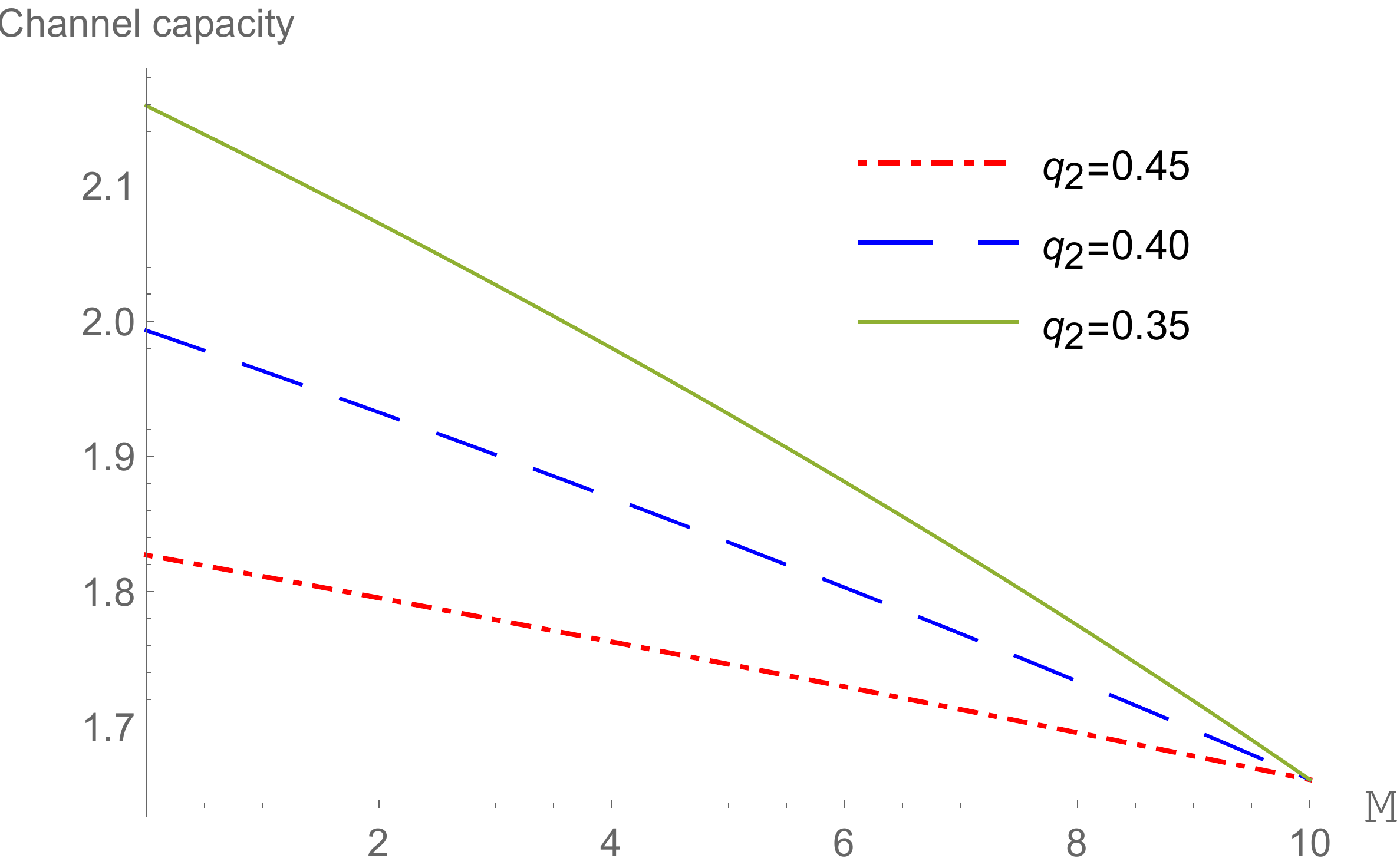}
\caption{Channel capacity as a function of $M$ for $N=10$, $q_{1}=0.5$ and three values of $q_{2}$.} 
\end{figure}
As expected, the channel capacity is lowest for high values of $q_{2}$ and low values of $M$.
\begin{figure}
\label{Fig1}
\includegraphics[scale=.4]{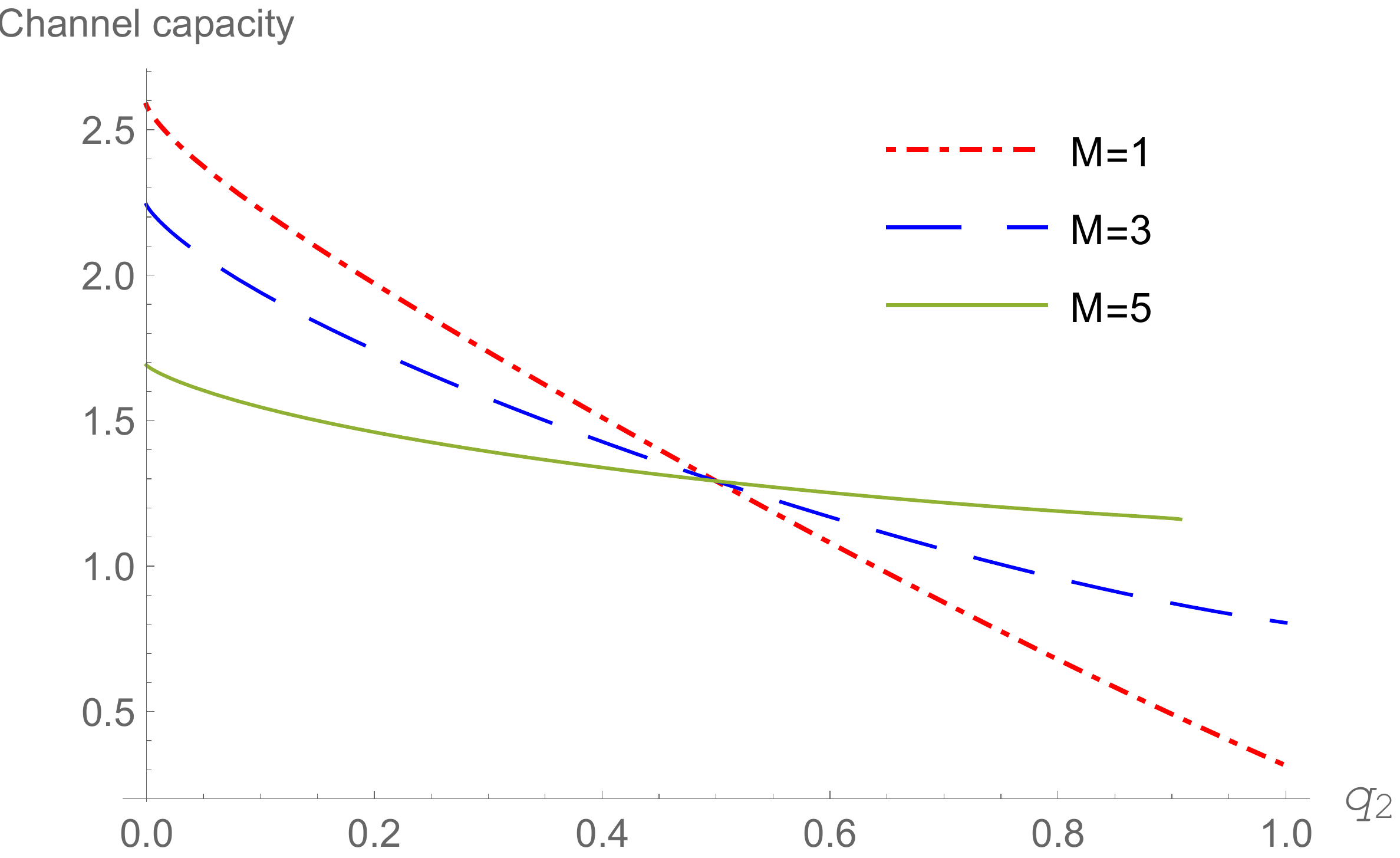}
\caption{Channel capacity as a function of $q_{2}$ for $N=6$, $q_{1}=0.5$, and three different values of $M$. } 
\end{figure} 

\section{A simple eavesdropping attack}
The scheme we have proposed, successive unambiguous discrimination measurements on the same qudit, can be useful for constructing a quantum communication protocol that uses nonorthogonal quantum states.  We would like now to investigate the robustness of this scheme against a simple eavesdropping attack, for the equal overlap case, and see how the information gained by the eavesdropper depends on the dimension of the transmitted system.  If Eve captures the qudit that was sent to one of the parties,  she may choose to apply unambiguous state discrimination to the qudit, but there is a probability that her measurement will fail. If it does, she will have to guess the state that needs to be sent to the next party.  This means that she will gain no information about the qudit and will introduce errors.  In the case of qubits, a better strategy is for Eve to use minimum-error state discrimination to try to identify the state \cite{bergou2}.  Unlike unambiguous discrimination, minimum-error discrimination returns a result every time, but the result can be incorrect.  The probability of making an error, however, is minimized. 

For symmetric states, the minimum-error measurement is known \cite{ban,eldar}.  The minimum-error POVM elements for the qudit states $\{|\eta_{j}\rangle\}$ for $ j=1,\ldots, N$  are given by
\begin{equation}
\label{min-error}
\Pi_{j}=\frac{1}{N} \rho^{-1/2}|\eta_{j}\rangle\langle\eta_{j}|\rho^{-1/2} ,
\end{equation}
where $\rho = (1/N)\sum_{j=1}^{N} |\eta_{j}\rangle\langle\eta_{j}|$.  The POVM elements and the success probability for the equal-overlap case were derived by Englert and \v{R}eha\v{c}ek \cite{englert}.  For the sake of completeness, we work them out in our notation in an appendix. 
If each state is equally likely, the probability that Eve successfully identifies the state is given by:
\begin{multline}
P^{(Eve)}_{success}  = \frac{1}{N}\sum_{j=1}^{N} \langle \eta_{j}|\Pi_{j}|\eta_{j}\rangle\\
  =  \frac{1}{N^{2}}\left( \sqrt{1+(N-1)s}+(N-1)\sqrt{1-s} \right)^{2}
\end{multline}
Figure 1 shows Eve's success probability as a function of $s$ for different values of $N$. We note that  as $N$ increases, the success probability decreases and approaches $1-s$ as a limit.  Therefore, using qudits rather than qubits, the success probability of someone using a minimum-error strategy to determine the transmitted state is decreased.  Assuming that Eve sends a copy of the same state she detects, this would also lead to an increased error probability for the legitimate users. They can detect these by comparing publicly a subset of their results.  If there are errors, then an eavesdropper was present.

\begin{figure}
\label{Fig3}
\includegraphics[scale=.35]{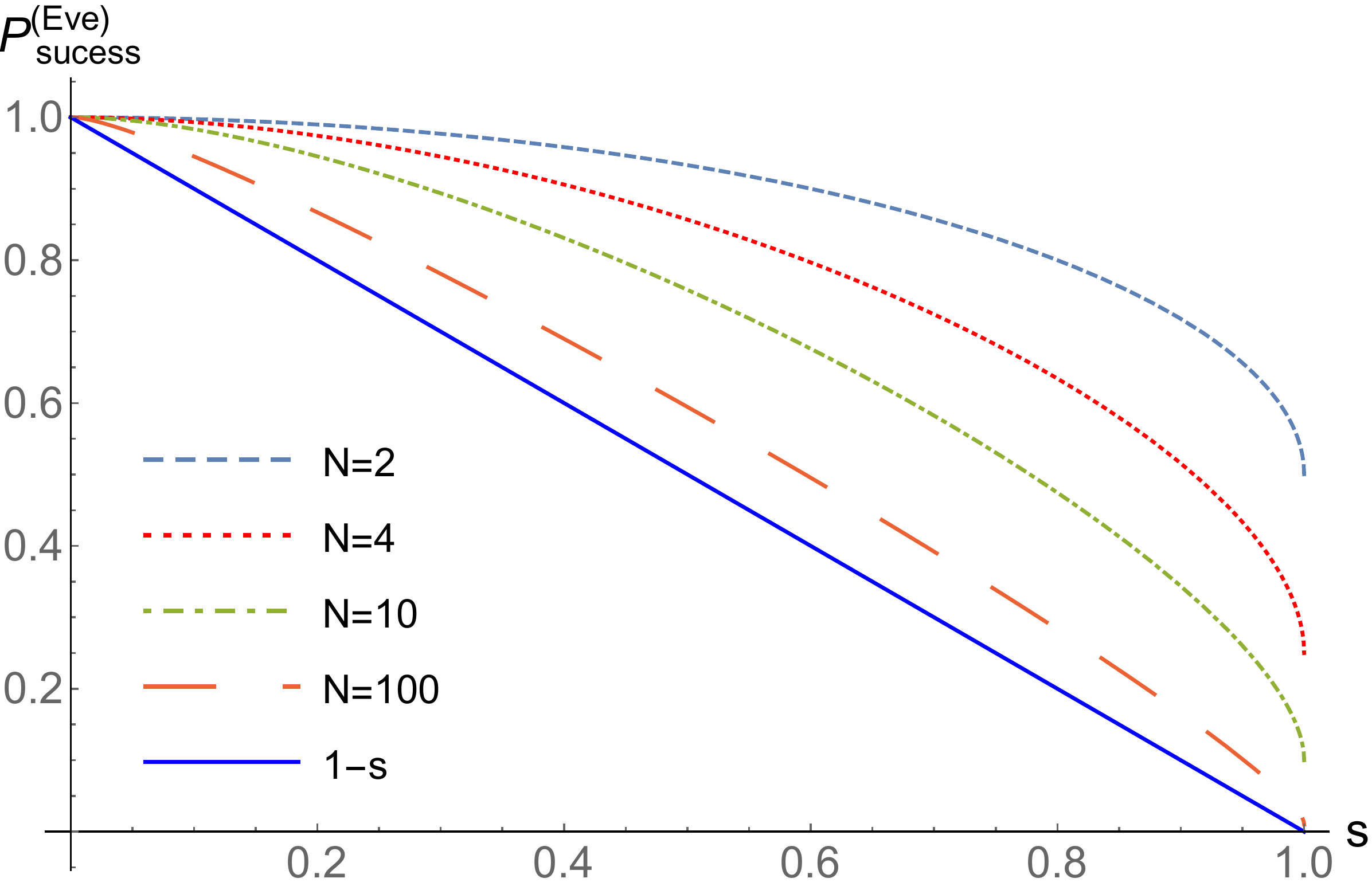}
\caption{Eve's success probability as a function of $s$ for different values of $N$.  As $N$ increases, the success probability goes to $1-s$.} 
\end{figure}

\section{Conclusion}
We have studied a scheme in which successive unambiguous discrimination measurements are made on qudits in order to determine which state was originally sent.  Two different scenarios were studied, the first in which all of the states have the same overlap and the second, in which there are two sets of states, with the states within each set having the same overlap, but the overlaps being different for the two sets.  The methods developed here can be extended to the case of more sets with different overlaps.  The results were compared to those in which qubits are sent in order to determine the effect of using a higher dimensional system.  We found that the failure probabilities of the unambiguous discrimination measurements for the case of N qudits of dimension $N$ are essentially the same, for the examples we examined, as they are for two qubit states.  This means that there is no penalty for using qudits, and therefore, qudits can carry more information per transmission than qubits, and we explicitly calculated channel capacities to show this.  We also studied a simple eavesdropping scheme and found that the use of qudits leads to a greater probability of error for the eavesdropper and, thereby, a greater probability that she will introduce errors that will be detected by legitimate users.

\section*{Acknowledgment}
MH was supported by a grant from the John Templeton Foundation. JM acknowledges support from a Naval Innovative Science and Engineering (NISE) Basic and Applied Research grant.

\section*{Appendix A}
We wish to find the vector $|\eta_{j}^{\perp}\rangle$ that satisfies $\langle \eta_{j}^{\perp}|\eta_{k}\rangle =0$ for $j\neq k$ for the case in which the vectors are divided into two sets, with different overlaps for the different sets.  As before we expand $|\eta_{j}^{\perp}\rangle$ in terms of the vectors $|\eta_{k}\rangle$
\begin{equation}
|\eta_{j}^{\perp}\rangle = \sum_{l=1}^{N} d_{l} |\eta_{l}\rangle ,
\end{equation}
and define the quantities
\begin{eqnarray}
\Lambda_{1}& = &  \sum_{l=1}^{M} d_{l} \nonumber \\
\Lambda_{2} & = & \sum_{l=M+1}^{N} d_{l} .
\end{eqnarray}
The condition $\langle \eta_{k}|\eta_{j}^{\perp}\rangle =0$ for $k\neq j$ gives us that for $k\neq j$ and $k\leq M$,
\begin{equation}
d_{k}= \frac{-1}{1-s_{1}^{2}} (s_{1}^{2} \Lambda_{1} + s_{1}s_{2}\Lambda_{2}) ,
\end{equation}
and for $k\neq j$ and $k\geq M+1$ 
\begin{equation}
d_{k} = \frac{-1}{1-s_{2}^{2}} (s_{1}s_{2}\Lambda_{1} + s_{2}^{2}\Lambda_{2}) .
\end{equation}

In order to find $d_{j}$, we have to consider two cases, $d_{j}\leq M$ and $d_{j}\geq M+1$.  We will first do the $j\leq M$ case and give the result for the $j\geq M+1$ case.  Summing the first if the above equations from $1$ to $M$ and the second from $M+1$ to $N$ we find
\begin{eqnarray}
\Lambda_{1}-d_{j} & = & \frac{-(M-1)}{1-s_{1}^{2}} (s_{1}^{2} \Lambda_{1} + s_{1}s_{2}\Lambda_{2})
\nonumber \\
\Lambda_{2} & = &  \frac{-(N-M)}{1-s_{2}^{2}} (s_{1}s_{2}\Lambda_{1} + s_{2}^{2}\Lambda_{2}) .
\end{eqnarray}
These equations allow us to find $\Lambda_{1}$ and $\Lambda_{2}$ in terms of $d_{j}$, which, in turn, allows us to find $d_{k}$ for $k\neq j$ in terms of $d_{j}$.  Defining
\begin{equation}
D_{1}  =  [1+s_{2}^{2}(N-M-1)](1-s_{1}^{2}) +(M-1)s_{1}^{2}( 1-s_{2}^{2}) ,
\end{equation}
we have that for $k\neq j$ and $k\leq M$
\begin{equation}
d_{k}=\frac{-s_{1}^{2}(1-s_{2}^{2})}{D_{1}} d_{j} ,
\end{equation}
and for $k\geq M+1$
\begin{equation}
d_{k}= \frac{-s_{1}s_{2}(1-s_{1}^{2})}{D_{1}} d_{j} 
\end{equation}
The normalization condition on the state then gives us
\begin{equation}
d_{j} = \left[ \frac{D_{1}}{(1-s_{1}^{2})(D_{1}+s_{1}^{2}(1-s_{2}^{2}))}\right]^{1/2} .
\end{equation}
This then gives us that for $jj\leq M$
\begin{equation}
\langle\eta_{j}|\eta_{j}^{\perp}\rangle = \frac{1}{D_{1}^{1/2}} [(1-s_{1}^{2})(D_{1}+s_{1}^{2}(1-s_{2}^{2}))]^{1/2} .
\end{equation}

Now let us look at the $j\geq M+1$ case.  Defining
\begin{equation}
D_{2}=s_{2}^{2}(1-s_{1}^{2})(N-M-1)+(1-s_{2}^{2})[1+s_{1}^{2}(M-1)]
\end{equation}
we find that for $k\leq M$
\begin{equation}
d_{k}= \frac{-s_{1}s_{2}(1-s_{2}^{2})}{D_{2}}d_{j}
\end{equation}
and for $k\geq M+1$ and $k\neq j$
\begin{equation}
d_{k}=\frac{-s_{2}^{2}(1-s_{2}^{2})}{D_{2}} d_{j} .
\end{equation}
From this we have that
\begin{equation}
d_{j} \left[ \frac{D_{2}}{(1-s_{2}^{2})(D_{2}+s_{2}^{2}(1-s_{1}^{2}))}\right]^{1/2} .
\end{equation}
Finally, we find that for $j\geq M+1$
\begin{equation}
\langle\eta_{j}|\eta_{j}^{\perp}\rangle = \frac{1}{D_{2}^{1/2}} [(1-s_{2}^{2})(D_{2}+s_{2}^{2}(1-s_{1}^{2}))]^{1/2} .
\end{equation}

\section*{Appendix B}
Here we derive the POVM elements for the minimum-error measurement for the equal-overlap case using our notation.  In order to make use of Eq.\ (\ref{min-error}), we have to diagonalize $\rho$ in order to find $\rho^{-1/2}$.  The eigenvalue equation $\rho |\Psi\rangle = \lambda |\Psi\rangle$ implies, if we set $|\Psi\rangle = \sum_{j=1}^{N}c_{j}|\eta_{j}\rangle$, that
\begin{equation}
\frac{1}{N} \sum_{j=1}^{N} [c_{j}+ s(\sum_{k\neq j} c_{k})] |\eta_{j}\rangle = \lambda \sum_{j=1}^{N} c_{j}|\eta_{j}\rangle .
\end{equation}
Defining $\Gamma = \sum_{j=1}^{N}c_{j}$, this implies, since the $|\eta_{j}\rangle$ are linearly independent, 
\begin{equation}
c_{j}+s(\Gamma - c_{j}) = Nc_{j} ,
\end{equation}
or
\begin{equation}
c_{j}=\frac{s\Gamma}{N\lambda + s -1} .
\end{equation}
Summing this over $j$ we obtain the consistency condition
\begin{equation}
\Gamma = \frac{Ns\Gamma}{N\lambda + s-1} .
\end{equation}
There are two cases, $\Gamma \neq 0$ and $\Gamma = 0$.  In the first case, we find that all of the $c_{j}$'s are equal, and 
\begin{equation}
\lambda = \frac{1}{N} [1+(N-1)s] .
\end{equation}
The corresponding normalized eigenvector is
\begin{equation}
|u_{1}\rangle = \frac{1}{\sqrt{N} [1+(N-1)s]^{1/2}}\sum_{j=1}^{N} |\eta_{j}\rangle .
\end{equation}
If $\Gamma = 0$, then the eigenvalue condition becomes
\begin{equation}
\lambda = \frac{1}{N}(1-s) ,
\end{equation}
because at least one of the $c_{j}$'s must be nonzero.  This eigenvalue is $(N-1)$-fold degenerate.  Therefore,
\begin{equation}
\rho = \frac{1}{N} [1+(N-1)s] |u_{1}\rangle\langle u_{1}| + \frac{1}{N}(1-s) (I-|u_{1}\rangle\langle u_{1}|) ,
\end{equation}
and 
\begin{eqnarray}
\rho^{-1/2} & = & \left[ \frac{N}{1+(N-1)s}\right]^{1/2} |u_{1}\rangle\langle u_{1}| \nonumber \\
 & & + \left( \frac{N}{1-s}\right)^{1/2}(I-|u_{1}\rangle\langle u_{1}|).
\end{eqnarray}
We find that 
\begin{eqnarray}
\rho^{-1/2} |\eta_{j}\rangle & = & \left[ 1-\left(\frac{1+(N-1)s}{1-s}\right)^{1/2}\right] |u_{1}\rangle \nonumber \\
& & + \left(\frac{N}{1-s}\right)^{1/2}|\eta_{j}\rangle .
\end{eqnarray}
The norm of this vector is found to be $\sqrt{N}$, so this implies that the vector
\begin{equation}
|\gamma_{j}\rangle = \frac{1}{\sqrt{N}} \rho^{-1/2} |\eta_{j}\rangle ,
\end{equation}
is normalized.  Furthermore, since $\Pi_{j}=|\gamma_{j}\rangle\langle \gamma_{j}|$, the minimum-error measurement is just a projective measurement.

\end{document}